# Two-Dimensional Ferromagnetism in Monolayers of MnSi

*Yuan Fang,[a] Yang Liu,[a] Dmitry V. Averyanov,[b] Ivan S. Sokolov,[b] Alexander N. Taldenkov,[b] Oleg E. Parfenov,[b] Oleg A. Kondratev,[b] Andrey M. Tokmachev,[b] Vyacheslav G. Storchak*,[b]*

[a] Center for Correlated Matter and School of Physics, Zhejiang University, Hangzhou 310058, China

[b] National Research Center "Kurchatov Institute", Kurchatov Sq. 1, 123182 Moscow, Russia

E-mail: vgstorchak9@gmail.com

Keywords: MnSi, 2D ferromagnetism, monolayer, metal-insulator transition, Kondo effect

2D ferromagnets offer valuable insights into the fundamentals of magnetism and stimulate the progress of ultracompact spintronics. The demand for seamless integration of the materials with the Si technology, particularly helpful to their applications in nanoelectronics, draws attention to 2D magnetic silicides. MnSi is a prominent silicide hosting magnetic phases with unconventional properties; however, little is known about magnetic states of MnSi at the 2D limit. Here, we explore the magnetism of ultrathin films of MnSi on silicon, down to a single monolayer. Angle-resolved photoemission spectra suggest exchange splitting of MnSi bands. Magnetization measurements confirm that the ferromagnetic state in MnSi is rather robust with respect to the number of monolayers. Thick metallic films demonstrate the anomalous Hall effect and negative magnetoresistance; however, as the number of monolayers drops below 3, MnSi becomes an insulator. Most importantly, the ferromagnetism of ultrathin MnSi films acquires a 2D character, as its effective Curie temperature depends on weak magnetic fields. The present study establishes MnSi monolayers as 2D ferromagnets that can find potential applications in silicon-based spintronics.

## 1. Introduction

Demands for ultracompact spintronics and advances in synthesis of 2D materials drive research on the 2D limit of magnetism.[1,2] 2D magnets are suggested as a source of unconventional magnetic phenomena and quantum phases.[3] They promise a wide range of applications in spintronics[4] and opto-spintronics,[5] especially as functional layers in complex heterostructures.[6] The advanced properties of 2D magnets stem from their high amenability to external stimuli[7] such as magnetic fields,[8] gating[9] or strains.[10] Coupling of magnetism to charge transport produces colossal magnetoresistance effects.[11,12] Research on 2D magnets provides insights into the fundamentals of magnetism, including spin dimensionality,[13] magnetic anisotropy,[14] spin dynamics[15] or magnon physics.[16] The bulk of studies on 2D magnets employ van der Waals (vdW) materials because they can be readily exfoliated and incorporated into vdW heterostructures. However, research on non-vdW 2D magnets[17,18] is also gaining momentum. The magnetic properties of such materials are sensitive to the reduced dimensionality; in the 2D limit, the magnetic state may undergo radical transformations[19,20] or the pre-existing magnetism of a material may become enhanced;[21] the magnetic state may be controlled chemically.[22] The emerging interest in 2D non-vdW materials facilitates studies of the 2D limit of unconventional magnets, in particular, altermagnets.[23] Non-vdW 2D magnets are expected to find various applications in nanoelectronics; they can be used as spintronic materials,[24,25] interconnects[26] or photocatalysts.[27] An important group of non-vdW 2D magnets is silicides;[19,23,26,28] seamless integration with the ubiquitous Si technology gives them a crucial advantage. Therefore, a coherent route to new functional non-vdW 2D magnets would be to take a prominent member of the silicide family and bring it to the 2D limit.

Among magnetic silicides, special attention is paid to MnSi, a paradigmatic mixed-valence metallic compound that has long been known for its unconventional properties. In its bulk form, it is an itinerant magnet with a Curie temperature of about 29 K, suppressed by applying a pressure of 1.4 GPa. One of the most prominent features of MnSi is its extended non-Fermi-liquid phase[29] exhibiting an unusual pattern of magnetic moments with partial long-range order.[30] The magnetic phase diagram of MnSi is rather complex.[31] In the magnetically ordered phase, the moments form an incommensurate spiral with a period of 18 nm in the [111] direction of the B20 crystal structure. Application of a magnetic field transforms the helimagnetic ground state into a conical structure; at still higher fields, MnSi enters into a field induced ferromagnetic (FM) state. The FM moments at saturation amount to 0.4 $\mu_B$/Mn, well below the Curie-Weiss magnetic moments of 2.2 $\mu_B$/Mn. Various magnetic species comprising several Mn atoms have been detected in MnSi.[32,33] Mostly, MnSi is renowned for the presence of skyrmions in the so-called *A* phase appearing on its phase diagram near $T_C$,[34,35] which may lead to applications in spintronics.[36] It is only natural to wonder how such an unusual magnetic material as MnSi would behave at the 2D limit.

Most 2D magnets are insulators; only a few are considered metallic.[9,26,37] However, it is common for these magnets to lose their metallic properties at the 2D limit: for instance, Fe$_3$GeTe$_2$, the best known metallic 2D magnet, becomes increasingly more insulating as its thickness decreases.[9] Therefore, the metal-insulator transition in MnSi is of particular interest.

Production of nanostructured MnSi materials has drawn significant attention; it applies to both 1D[38,39] and 2D[40,41] materials. In thin films, their synthesis conditions,[42-44] atomic[45,46] and electronic[41] structure as well as magnetic properties[40,47-49] have been studied extensively. In general, as long as the film is not ultrathin, it exhibits magnetic and electronic properties similar to those of bulk MnSi.[47] However, there are important differences: in thin films, (i) the non-Fermi-liquid behavior is observed in a wider temperature range though requires higher pressures;[50] (ii) the range of temperatures and magnetic fields for the skyrmionic phase is also extended;[51,52] (iii) the period of the spin helix is almost twice as short;[51] (iv) Mn$^+$ oxidation state is stabilized;[53] (v) the Curie temperature is enhanced significantly, to 43-45 K,[40,41,45,48-50] often attributed to strain effects in ultrathin films. The origin of these differences is controversial, especially regarding the enhanced $T_C$.[40,49] It should be stressed that the focus of studies on the magnetic structure of MnSi has been on the bulk or rather thick films. Accordingly, little is known about the 2D limit of MnSi magnetism, especially at the level of monolayers (ML).

Here, we fill this gap by analyzing MnSi films with a thickness down to a single ML. We report their synthesis, structural characterization by diffraction techniques, band structure studies using angle-resolved photoemission spectroscopy (ARPES), measurements of magnetic and electron transport properties. We demonstrate the emergence of the metal-insulator transition and evolution of MnSi magnetism with the number of MLs. In particular, we observe a pronounced dependence of the effective Curie temperature on weak magnetic fields, a hallmark of 2D magnetism.

## 2. Results and Discussion

### 2.1. Synthesis and Characterization

Synthesis of MnSi films on silicon has been explored in a number of studies;[41-44] alternative approaches have been developed and tested. In particular, different techniques have been used, such as molecular beam epitaxy (MBE)[40,41,54,55] and magnetron sputtering.[44] MnSi can be grown on two faces of the Si substrate, namely (001)[44,55] and (111).[40-43,54] MBE synthesis of a MnSi film can be carried out using two alternative procedures: one is based on co-deposition of Mn and Si,[42,48] the other on deposition of Mn alone with subsequent annealing.[40-43,54] In the latter case, the substrate serves as the source of the silicon necessary for reaction. MBE synthesis on Si(111) using reaction of Mn with the substrate is probably the most developed route to MnSi films. It results in homogeneous films with atomically flat interfaces;[56] the procedure provides a coherent epitaxy of ultrathin films of MnSi.[57] The main type of defects in such films, holes or craters,[58] is unlikely to make a strong impact on the magnetic properties. A similar synthetic route based on reaction of a metal with the Si(111) substrate has been successfully employed for other 2D magnetic silicides as well.[12,19,26,28] Therefore, it is chosen here to produce the MnSi films for our studies.

The synthesis of the films starts with preparation of the substrate surface. The Si wafer is covered with an oxide layer; the oxide is removed thermally under high vacuum. The formation of the bare Si(111) surface is witnessed *in situ* using reflection high-energy electron diffraction (RHEED) – the oxide removal exposes the typical $7 \times 7$ reconstruction of Si(111). The synthetic procedure depends on the intended thickness of the film. Although MnSi is not a layered compound, this thickness is sometimes measured in terms of MLs,[40,41] defined as the basic repeating unit along the [111] direction comprising 2 layers of Mn and 2 layers of Si (and thus also called a quadruple layer[54,57]). The films with a thickness of up to 3 ML are produced by deposition of the corresponding amounts of Mn on top of Si(111) and subsequent annealing (see details in the Methods section). To produce thicker films, a 3 ML layer of MnSi is grown first; then, the rest is synthesized directly by deposition of Mn at high temperature. It is presumed that the Si atoms from the substrate diffuse to the surface of the film *via* the vacancy mechanism. In general, the synthesis is carried out according to a procedure reported in Ref. [41]. The films are naturally integrated with the silicon substrate. To protect them from oxidation by air, they are capped with a layer of amorphous Si. The capping layer does not affect the results of structural characterization as well as measurements of magnetic and transport properties of the films. In total, 4 ultrathin films of MnSi are selected for further experiments – 1 ML, 2 ML, 3 ML, and 5 ML. Also, a film with a thickness of 20 nm is synthesized for comparison with other works on MnSi.

The atomic structure has been studied using diffraction techniques. In particular, RHEED, a technique sensitive to the film surface, has been instrumental for *in situ* control of the synthesis. Figure 1 demonstrates characteristic RHEED patterns for ultrathin films of MnSi, 1 ML to 3 ML. Basically, they show the same structure of reflexes corresponding to epitaxial MnSi films on Si(111); they are quite similar to the RHEED image reported in Ref. [41] for a thicker film. The RHEED patterns are sharp and exhibit a high contrast which is an indicator of the high quality of the films. The extracted lateral lattice constant is 3.33(1) Å. After capping, the atomic structure of the films was probed using X-ray diffraction (XRD). The resulting θ-2θ scans are presented in Figure 2. As expected, the MnSi peaks become more pronounced as the film thickness increases. The main conclusions are the following: MnSi grows epitaxially on Si(111); the side phases are absent. The thickness of one ML, deduced from the XRD peak positions for 20 nm MnSi, is 2.626(2) Å; this value can be compared with that of 2.624 Å reported in Ref. [41].

To study the electronic structure of the ultrathin films in the FM state, ARPES measurements were carried out at $T = 6$ K, well below $T_C$ (see below). The data for 1-3 ML are presented in Figure 3. The spectra for 2 and 3 ML MnSi are very similar to those for thicker MnSi films:[41] they both show parabolic bands near -1.0 eV around $\overline{\Gamma}$ and dispersive flat bands near the Fermi level. According to the previous study,[41] the ferromagnetism in MnSi films is manifested by exchange splitting of ~0.2 eV in the electronic bands, which is best seen near $k_y \sim 1.12$ Å$^{-1}$. Figures 3d-f show the energy distribution curves (EDC) extracted from ARPES spectra in Figures 3a-c, which reveal two split peaks with an energy separation of ~0.22 eV for the 2-3 ML films (highlighted by two arrows in Figures 3e,f). Such splitting vanishes for $T > T_C$[41] – this observation points at ferromagnetism in the 2-3 ML films. For the 1 ML film, the main band structure remains somewhat similar to those of 2 and 3 ML films (Figure 3a), although there is a new flat band near -0.8 eV, which origin is unclear. At the same time, the exchange splitting becomes difficult to identify due to the broad peak at -0.3 eV (Figure 3d). This might be caused by possible disorder at the interface, which makes the spectrum somewhat blurry. All in all, the observation of exchange splitting calls for magnetization measurements in ultrathin films.

## 2.2. Magnetism of MnSi Films

The MBE grown MnSi films have a large area; therefore, despite their extreme thinness, magnetic properties of MnSi MLs can be studied using SQUID. Thick MnSi films are known to exhibit easy-plane anisotropy;[56] the same anisotropy is exhibited by ultrathin films (see Figure S1 of the Supporting Information). Accordingly, the magnetization measurements discussed below are carried out in in-plane magnetic fields. We are interested in ultrathin films but it is worthwhile to make an additional check of magnetic properties of MnSi in a rather thick film to make a comparison with the literature data. Figure S2 of the Supporting Information demonstrates temperature dependence of the FM moment in 20 nm MnSi. The Curie temperature in thick MnSi films can be derived from temperature dependent magnetization by taking a derivative[49] or using a 4-parameter fit;[48] the former approach is simpler and does not require a model of the FM transition. The derivative of the FM moment with respect to temperature (Figure S2 of the Supporting Information) determines $T_C$ to be about 43 K. This value is in agreement with a number of other works.[40,41,45,48-50]

Despite the large area, the measurements of monolayer samples are still challenging as the signals are too low. Therefore, we have chosen 3 ML MnSi as the main object for illustration of magnetic properties – it is sufficiently thin to exhibit 2D magnetism yet the magnetic signals are not too noisy as in the case of 1 ML. First, we give a general outlook of magnetic properties of ultrathin MnSi films. Figure 4 provides this information for 3 ML MnSi whereas Figures S3-S5 of the Supporting Information – for 5 ML, 2 ML, and 1 ML MnSi, respectively. Figure 4a shows magnetic field dependence of the FM moment, which demonstrates a distinct hysteresis loop. In 2D magnets, the hysteresis loop usually shrinks[9] and sometimes disappears altogether.[59] This behavior is not related to the magnetic moment – for instance, a pronounced hysteresis loop in 2 ML EuGe$_2$ disappears in 1 ML EuGe$_2$ although the magnetic moment in the latter is 2.5 times higher.[20] In 1 ML MnSi, a hysteresis loop is still discernible although the magnetic signal is very noisy. Figure 4a shows that the saturation magnetic moment in 3 ML MnSi is about 0.4 $\mu_B$/Mn, *i.e.* basically of the same value as in bulk MnSi. The saturation magnetic moment in 1 ML MnSi, determined in our measurements, is about 0.35 $\mu_B$/Mn, *i.e.* it is rather stable with respect to the film thickness.

Figure 4b provides magnetic field dependence of the FM moment for a range of temperatures. It demonstrates that the moment gradually diminishes as the temperature rises and then disappears above $T_C$. The films exhibit other characteristics typical of FM materials. In particular, Figure 4c shows the remnant moment in 3 ML MnSi and its decay with temperature. At low temperature, field-cooled (FC) and zero-field-cooled (ZFC) magnetization curves diverge (Figure 4d). The observed pattern of the FC/ZFC curves is expected for ferromagnets exhibiting hysteresis.[60] The measurements are carried out using a magnetic field (7 mT) that is well below the coercive field $H_c$ at low temperatures; therefore, the random orientation of pinned FM domains results in a relatively low ZFC magnetic moment. At higher temperature, the magnetic field exceeds $H_c$: the FC and ZFC curves converge because the external magnetic field determines the magnetization dynamics in both cases.

In 2D magnets, important conclusions can be made based on temperature dependences of the FM moment in low magnetic fields. Figure 5 provides this information for 3 ML MnSi whereas Figures S6-S8 of the Supporting Information – for 5 ML, 2 ML, and 1 ML MnSi, respectively. Figure 5a, showing $M_{FM}(T)$ curves for 3 ML MnSi, signifies that the FM transition takes place at a rather high temperature which is not very much different from that in a thick film. This is in sharp contrast to previous works that suggested that $T_C$ is strongly reduced in ultrathin films:[48,55] for instance, a $T_C$ of 4 K is reported for 4 ML MnSi in Ref. [55]. The discrepancy may be due to differences in the quality of the MnSi films. Determination of $T_C$ in ultrathin MnSi is not trivial. Different techniques are not in agreement, in contrast to thick MnSi films.[40] Most significantly, the effective $T_C$ appears to depend on the magnetic field, as seen in Figure 5a. This fact is better visualized using plots of the normalized FM moment (see Figure 5b), where the shifts in the $M_{FM}(T)$ curves can be seen more clearly. Similar shifts are detected for 5 ML and 2 ML MnSi; the situation with 1 ML MnSi is less obvious because the data are too noisy at this featherweight level. The dependence of $T_C$ on weak magnetic fields in 2D magnets has been described theoretically[59,61] and witnessed experimentally in $Cr_2Ge_2Te_6$[59] and various rare-earth silicides and germanides.[19,20,28,62] This dependence is a fingerprint of 2D ferromagnetism. The phenomenon arises because of the influence of magnetic fields on the spin-wave excitation spectra in 2D magnets.[61] The spin-wave amplitude in an isotropic 2D system with short-range interactions diverges at any finite temperature. The long-range magnetic order is stabilized *via* opening a (pseudo)gap in the spectrum due to magnetic anisotropy or dipolar interaction (the latter is particularly important for materials with easy-plane magnetic anisotropy, like MnSi films). Weak magnetic fields control this (pseudo)gap[61] making the $M_{FM}(T)$ curves strongly dependent on the external field. This feature makes a sharp contrast with the standard 3D magnetism in which $T_C$ is robust with respect to external fields.

## 2.3. Electron Transport Properties

The magnetism in MnSi films affects electron transport; so, the latter can be used as an alternative way to probe magnetic properties in the 2D limit. MnSi is a metallic material, suggesting that lateral electron transport can be studied. Generally, the conductivity depends on the number on MLs $N$. It is typical for a metal to undergo a metal-insulator transition as $N$ diminishes, although some materials are capable to retain their metallic character down to a single ML.[26] Our study of the MnSi films reveals that the films with $N \geq 3$ are metallic whereas 1 ML and 2 ML MnSi are insulating. ARPES data suggest 1 ML and 2 ML films to be metallic. Such a discrepancy between electron transport and ARPES results are known for other 2D magnets.[9,63] Figure 6a demonstrates the dependence of the layer normalized sheet conductance on $N$ for the 3 metallic films. As the MnSi layer approaches the 2D limit, it becomes progressively less and less conducting. An $N$-dependent metal-insulator transition in MnSi has been reported earlier[64] though for a larger value of $N$ – the discrepancy may be caused by different amounts of defects in the films. The results of magnetization measurements provided above suggest that magnetic properties of MnSi are stable with respect to the metal-insulator transition.

The electron transport in the insulating 1 ML and 2 ML films is difficult to analyze because of the concurrent conductivity of the Si substrate. Therefore, we again employ 3 ML MnSi as the main system for analysis. Figure 6b shows temperature dependence of sheet resistivity which exhibits a feature marking the magnetic transition in MnSi. A typical behavior of low-temperature resistivity in rather thick MnSi films is its increase following the $T^2$ law.[65] In contrast, $R(T)$ in 3 ML MnSi decreases at low temperature; moreover, the decrease is proportional to $lnT$. A logarithmic temperature dependence of resistivity in MnSi has been reported in Ref. [64] and explained by weak anti-localization. In our experiments, 3 ML MnSi does not exhibit the strong anisotropy of MR characteristic to weak localization effects. In particular, Figure S9 of the Supporting Information demonstrates logarithmic $R(T)$ curves in various out-of-plane and in-plane magnetic fields. In Ref. [41], the discovered $lnT$ dependence in a study of ARPES intensity in a MnSi film has been associated with Kondo physics. In fact, the situation in 3 ML MnSi is quite similar to that in 4 ML GdSi$_2$, another magnetic silicide on the brink of the metal-insulator transition, which shows a logarithmic dependence of resistivity attributed to the Kondo interaction of carriers with local magnetic moments.[12]

Typically, the effect of the film magnetism on the electron transport is revealed in studies of MR and the anomalous Hall effect (AHE). Figure 6c demonstrates negative MR at low temperature in 3 ML MnSi, which can be associated with the FM order. Negative MR is observed in thick MnSi films as well.[65] Figure 6d shows the AHE in 3 ML MnSi, an effect that is typical of FM materials. It demonstrates a significant hysteresis associated with the hysteresis loop in the magnetic field dependence of the magnetic moment in the MnSi film. The Hall effect in ultrathin MnSi differs from that in thicker films (and the bulk) – the topological Hall effect[44,51] is absent. This is not surprising because the film thickness is probably too small to accommodate skyrmions. Figure S10 of the Supporting Information demonstrates that the negative MR and the AHE are not limited to 3 ML but observed in 5 ML MnSi as well.

## 3. Conclusion

2D ferromagnets are heralded as materials that can offer new insights into magnetism and foster the development of next-generation spintronic devices. A key issue is that the list of available 2D magnets is rather short. It is important that in search for new 2D magnets we are not limited to vdW materials. Non-vdW 2D magnets may be more suitable for integration with the current technologies. In particular, silicides are natural candidates for applications in Si electronics. Here, we explored ML-thick films of MnSi to establish this material as a 2D magnet. MnSi has been a subject of numerous studies due to its extraordinary properties but its potential as a 2D magnet has been overlooked. We synthesized epitaxial films of MnSi down to a single ML and studied their atomic and electronic structure, magnetic and transport properties using a combination of techniques. Our ARPES data suggest exchange splitting of the bands; a robust FM state is evidenced by studies of temperature and magnetic field dependences of the FM moments, the remnant moments and FC/ZFC bifurcation, the AHE and negative MR in lateral electron transport. A particularly interesting feature is the dependence of the effective Curie temperature on weak magnetic fields. It signifies that the FM state in ultrathin MnSi is of 2D nature. Moreover, the transition to the ML limit does not lead to any significant reduction of the Curie temperature or the saturation magnetic moment. All things considered, the system MnSi/Si makes a prospective material for applications in spintronics, including Si spintronics. MnSi is just one representative of the extended group of magnetic silicides and germanides formed by transition metals. It is likely that using similar synthetic approaches they can be produced as epitaxial films on Si and Ge, respectively. Therefore, we envisage that the present study will seed a class of 2D magnets integrated with technologically important semiconductors.

## 4. Methods

*Synthesis*: The MnSi films were synthesized under ultra-high vacuum conditions employing an MBE system; the base pressure in the growth chamber was kept below $2 \cdot 10^{-10}$ mbar. The Si(111) substrates, wafers of a typical size 5 mm × 10 mm, were heated to 900-1000 °C to remove the natural surface oxide. The 7 × 7 reconstruction of the bare Si(111) surface was verified by RHEED. MnSi was synthesized by reaction of Mn with the substrate. Mn deposition with a rate of about 0.7 ML min$^{-1}$ was carried out by heating an effusion cell to 790 °C; the amount of deposited Mn was controlled by a quartz crystal monitor. Different synthesis protocols were used for ultrathin films, up to 3 ML, and thicker films. Layers of Mn with thicknesses of 1.4 Å, 2.8 Å, and 4.2 Å were deposited on bare Si(111) kept at 70 °C and then annealed at 260 °C to form 1 ML, 2 ML, and 3 ML MnSi films, respectively. Thicker films employed 3 ML MnSi as a seeding layer; the rest of MnSi was synthesized by deposition of Mn on top of the film at 260 °C. The Si for the synthesis was supplied by diffusion from the substrate to the surface. To protect the films from air, they were capped at room temperature with a 10-nm layer of amorphous Si.

*Characterization*: The structural quality of the films was monitored *in situ* by RHEED using an electron gun operating at an accelerating voltage of 15 kV. XRD studies of MnSi were carried out in a Rigaku SmartLab 9 kW diffractometer employing a $CuK_{\alpha 1}$ radiation source (the wavelength is 1.54056 Å). The ARPES measurements were taken in a Helium-lamp ARPES system connected under ultra-high vacuum to the MBE system. The base pressure of the ARPES system was $7.6 \cdot 10^{-11}$ mbar which increased to $2.5 \cdot 10^{-10}$ mbar during the Helium lamp operation. A five-axis manipulator cooled by a closed-cycle helium refrigerator was employed for sample orientation and temperature control during the ARPES measurements. The photon source was a VUV-5k Helium lamp coupled to a grating monochromator. The measurements, carried out at 6 K, employed He-I (21.2 eV) photons. The energy and momentum resolutions were 12 meV and 0.01 Å$^{-1}$, respectively. The magnetic properties of MnSi films were probed by an MPMS XL-7 SQUID magnetometer employing the reciprocating sample option. Square samples 5 mm × 5 mm were oriented with respect to external magnetic field with accuracy better than 2°. The subtraction of the diamagnetic contribution from the Si substrate was carried out following the recipe of Ref. [19]. The lateral electron transport in MnSi films was measured by a Lake Shore 9709A system using square samples with a lateral size of 5 mm. The electrical contacts for four-contact measurements (adhering to the ASTM Standard F76) were produced by deposition of an Ag-Sn-Ga alloy; I-V characteristic curves attested the contact ohmicity.

**Acknowledgements**

This work was supported by NRC "Kurchatov Institute". The measurements were carried out using equipment of the resource centers of electrophysical and laboratory X-ray techniques at NRC "Kurchatov Institute". YL is supported by the National Key R&D Program of China (Grants No. 2023YFA1406303, No. 2022YFA1402200).

**References**

[1]    B. Huang, M. A. McGuire, A. F. May, D. Xiao, P. Jarillo-Herrero, X. Xu, *Nat. Mater.* **2020**, *19*, 1276.

[2]    S. Xing, J. Zhou, X. Zhang, S. Elliott, Z. Sun, *Progr. Mater. Sci.* **2023**, *132*, 101036.

[3]    K. S. Burch, D. Mandrus, J.-G. Park, *Nature* **2018**, *563*, 47.

[4]   Z. Jia, M. Zhao, Q. Chen, Y. Tian, L. Liu, F. Zhang, D. Zhang, Y. Ji, B. Camargo, K. Ye, R. Sun, Z. Wang, Y. Jiang, *ACS Nano* **2025**, *19*, 9452.

[5]   S. Liu, I. A. Malik, V. L. Zhang, T. Yu, *Adv. Mater.* **2025**, *37*, 2306920.

[6]   Z. Zhang, R. Sun, Z. Wang, *ACS Nano* **2025**, *19*, 187.

[7]   K. F. Mak, J. Shan, D. C. Ralph, *Nat. Rev. Phys.* **2019**, *1*, 646.

[8]   B. Huang, G. Clark, E. Navarro-Moratalla, D. R. Klein, R. Cheng, K. L. Seyler, D. Zhong, E. Schmidgall, M. A. McGuire, D. H. Cobden, W. Yao, D. Xiao, P. Jarillo-Herrero, X. Xu, *Nature* **2017**, *546*, 270.

[9]   Y. Deng, Y. Yu, Y. Song, J. Zhang, N. Z. Wang, Z. Sun, Y. Yi, Y. Z. Wu, S. Wu, J. Zhu, J. Wang, X. H. Chen, Y. Zhang, *Nature* **2018**, *563*, 94.

[10]  T. Song, Z. Fei, M. Yankowitz, Z. Lin, Q. Jiang, K. Hwangbo, Q. Zhang, B. Sun, T. Taniguchi, K. Watanabe, M. A. McGuire, D. Graf, T. Cao, J.-H. Chu, D. H. Cobden, C. R. Dean, D. Xiao, X. Xu, *Nat. Mater.* **2019**, *18*, 1298.

[11]  D. R. Klein, D. MacNeill, J. L. Lado, D. Soriano, E. Navarro-Moratalla, K. Watanabe, T. Taniguchi, S. Manni, P. Canfield, J. Fernández-Rossier, P. Jarillo-Herrero, *Science* **2018**, *360*, 1218.

[12]  O. E. Parfenov, A. M. Tokmachev, D. V. Averyanov, I. A. Karateev, I. S. Sokolov, A. N. Taldenkov, V. G. Storchak, *Mater. Today* **2019**, *29*, 20.

[13]  M. Gibertini, M. Koperski, A. F. Morpurgo, K. S. Novoselov, *Nat. Nanotechnol.* **2019**, *14*, 408.

[14]  A. Bedoya-Pinto, J.-R. Ji, A. K. Pandeya, P. Gargiani, M. Valvidares, P. Sessi, J. M. Taylor, F. Radu, K. Chang, S. S. P. Parkin, *Science* **2021**, *374*, 616.

[15]  C. Tang, L. Alahmed, M. Mahdi, Y. Xiong, J. Inman, N. J. McLaughlin, C. Zollitsch, T. H. Kim, C. R. Du, H. Kurebayashi, E. J. G. Santos, W. Zhang, P. Li, W. Jin, *Phys. Rep.* **2023**, *1032*, 1.

[16]  Y. J. Bae, J. Wang, A. Scheie, J. Xu, D. G. Chica, G. M. Diederich, J. Cenker, M. E. Ziebel, Y. Bai, H. Ren, C. R. Dean, M. Delor, X. Xu, X. Roy, A. D. Kent, X. Zhu, *Nature* **2022**, *609*, 282.

[17]  A. P. Balan, A. B. Puthirath, S. Roy, G. Costin, E. F. Oliveira, M. A. S. R. Saadi, V. Sreepal, R. Friedrich, P. Serles, A. Biswas, S. A. Iyengar, N. Chakingal, S. Bhattacharyya, S. K. Saju, S. C. Pardo, L. M. Sassi, T. Filleter, A. Krasheninnikov, D. S. Galvao, R. Vajtai, R. R. Nair, P. M. Ajayan, *Mater. Today* **2022**, *58*, 164.

[18]  H. Ren, G. Xiang, *Mater. Today. Electron.* **2023**, *6*, 100074.

[19]   A. M. Tokmachev, D. V. Averyanov, O. E. Parfenov, A. N. Taldenkov, I. A. Karateev, I. S. Sokolov, O. A. Kondratev, V. G. Storchak, *Nat. Commun.* **2018**, *9*, 1672.

[20]   A. M. Tokmachev, D. V. Averyanov, A. N. Taldenkov, O. E. Parfenov, I. A. Karateev, I. S. Sokolov, V. G. Storchak, *Mater. Horiz.* **2019**, *6*, 1488.

[21]   A. B. Puthirath, S. N. Shirodkar, G. Gao, F. C. R. Hernandez, L. Deng, R. Dahal, A. Apte, G. Costin, N. Chakingal, A. P. Balan, L. M. Sassi, C. S. Tiwary, R. Vajtai, C.-W. Chu, B. I. Yakobson, P. M. Ajayan, *Small* **2020**, *16*, 2004208.

[22]   T. Barnowsky, S. Curtarolo, A. V. Krasheninnikov, T. Heine, R. Friedrich, *Nano Lett.* **2024**, *24*, 3874.

[23]   O. E. Parfenov, D. V. Averyanov, I. S. Sokolov, A. N. Mihalyuk, O. A. Kondratev, A. N. Taldenkov, A. M. Tokmachev, V. G. Storchak, *J. Am. Chem. Soc.* **2025**, *147*, 5911.

[24]   M. Cheng, X. Zhao, Y. Zeng, P. Wang, Y. Wang, T. Wang, S. J. Pennycook, J. He, J. Shi, *ACS Nano* **2021**, *15*, 19089.

[25]   B. Das, S. Ghosh, S. Sengupta, P. Auban-Senzier, M. Monteverde, T. K. Dalui, T. Kundu, R. A. Saha, S. Maity, R. Paramanik, A. Ghosh, M. Palit, J. K. Bhattacharjee, R. Mondal, S. Datta, *Small* **2023**, *19*, 2302240.

[26]   O. E. Parfenov, D. V. Averyanov, I. S. Sokolov, A. N. Mihalyuk, O. A. Kondratev, A. N. Taldenkov, A. M. Tokmachev, V. G. Storchak, *Adv. Mater.* **2025**, *37*, 2412321.

[27]   A. P. Balan, S. Radhakrishnan, C. F. Woellner, S. K. Sinha, L. Deng, C. de los Reyes, B. M. Rao, M. Paulose, R. Neupane, A. Apte, V. Kochat, R. Vajtai, A. R. Harutyunyan, C.-W. Chu, G. Costin, D. S. Galvao, A. A. Martí, P. A. van Aken, O. K. Varghese, C. S. Tiwary, A. M. M. R. Iyer, P. M. Ajayan, *Nat. Nanotechnol.* **2018**, *13*, 602.

[28]   D. V. Averyanov, I. S. Sokolov, A. N. Taldenkov, O. E. Parfenov, O. A. Kondratev, A. M. Tokmachev, V. G. Storchak, *Small* **2024**, *20*, 2402189.

[29]   C. Pfleiderer, S. R. Julian, G. G. Lonzarich, *Nature* **2001**, *414*, 427.

[30]   C. Pfleiderer, D. Reznik, L. Pintschovius, H. v. Löhneysen, M. Garst, A. Rosch, *Nature* **2004**, *427*, 227.

[31]   A. Bauer, C. Pfleiderer, *Phys. Rev. B* **2012**, *85*, 214418.

[32]   V. G. Storchak, J. H. Brewer, R. L. Lichti, T. A. Lograsso, D. L. Schlagel, *Phys. Rev. B* **2011**, *83*, 140404(R).

[33]   Z. Jin, Y. Li, Z. Hu, B. Hu, Y. Liu, K. Iida, K. Kamazawa, M. B. Stone, A. I. Kolesnikov, D. L. Abernathy, X. Zhang, H. Chen, Y. Wang, C. Fang, B. Wu, I. A. Zaliznyak, J. M. Tranquada, Y. Li, *Sci. Adv.* **2023**, *9*, eadd5239.

[34] A. Neubauer, C. Pfleiderer, B. Binz, A. Rosch, R. Ritz, P. G. Niklowitz, P. Böni, *Phys. Rev. Lett.* **2009**, *102*, 186602.

[35] S. Mühlbauer, B. Binz, F. Jonietz, C. Pfleiderer, A. Rosch, A. Neubauer, R. Georgii, P. Böni, *Science* **2009**, *323*, 915.

[36] F. Jonietz, S. Mühlbauer, C. Pfleiderer, A. Neubauer, W. Münzer, A. Bauer, T. Adams, R. Georgii, P. Böni, R. A. Duine, K. Everschor, M. Garst, A. Rosch, *Science* **2010**, *330*, 1648.

[37] L. Meng, Z. Zhou, M. Xu, S. Yang, K. Si, L. Liu, X. Wang, H. Jiang, B. Li, P. Qin, P. Zhang, J. Wang, Z. Liu, P. Tang, Y. Ye, W. Zhou, L. Bao, H.-J. Gao, Y. Gong, *Nat. Commun.* **2021**, *12*, 809.

[38] K. Seo, H. Yoon, S.-W. Ryu, S. Lee, Y. Jo, M.-H. Jung, J. Kim, Y.-K. Choi, B. Kim, *ACS Nano* **2010**, *4*, 2569.

[39] H. Du, J. P. DeGrave, F. Xue, D. Liang, W. Ning, J. Yang, M. Tian, Y. Zhang, S. Jin, *Nano Lett.* **2014**, *14*, 2026.

[40] E. Karhu, S. Kahwaji, T. L. Monchesky, C. Parsons, M. D. Robertson, C. Maunders, *Phys. Rev. B* **2010**, *82*, 184417.

[41] Y. Fang, H. Zhang, D. Wang, G. Yang, Y. Wu, P. Li, Z. Xiao, T. Lin, H. Zheng, X.-L. Li, H.-H. Wang, F. Rodolakis, Y. Song, Y. Wang, C. Cao, Y. Liu, *Phys. Rev. B* **2022**, *106*, L161112.

[42] E. Magnano, E. Carleschi, A. Nicolaou, T. Pardini, M. Zangrando, F. Parmigiani, *Surf. Sci.* **2006**, *600*, 3932.

[43] S. Higashi, Y. Ikedo, P. Kocán, H. Tochihara, *Appl. Phys. Lett.* **2008**, *93*, 013104.

[44] W.-Y. Choi, H.-W. Bang, S.-H. Chun, S. Lee, M.-H. Jung, *Nanoscale Res.* **2021**, *16*, 7.

[45] A. I. Figueroa, S. L. Zhang, A. A. Baker, R. Chalasani, A. Kohn, S. C. Speller, D. Gianolio, C. Pfleiderer, G. van der Laan, T. Hesjedal, *Phys. Rev. B* **2016**, *94*, 174107.

[46] D. Morikawa, Y. Yamasaki, N. Kanazawa, Y. Yokouchi, Y. Tokura, T.-h. Arima, *Phys. Rev. Mater.* **2020**, *4*, 014407.

[47] E. Magnano, F. Bondino, C. Cepek, F. Parmigiani, M. C. Mozzati, *Appl. Phys. Lett.* **2010**, *96*, 152503.

[48] J. Engelke, T. Reimann, L. Hoffmann, S. Gass, D. Menzel, S. Süllow, *J. Phys. Soc. Jpn.* **2012**, *81*, 124709.

[49] Z. Li, Y. Yuan, V. Begeza, L. Rebohle, M. Helm, K. Nielsch, S. Prucnal, S. Zhou, *Sci. Rep.* **2022**, *12*, 16388.

[50] J. Engelke, D. Menzel, H. Hidaka, T. Seguchi, H. Amitsuka, *Phys. Rev. B* **2014**, *89*, 144413.

[51] Y. Li, N. Kanazawa, X. Z. Yu, A. Tsukazaki, M. Kawasaki, M. Ichikawa, X. F. Jin, F. Kagawa, Y. Tokura, *Phys. Rev. Lett.* **2013**, *110*, 117202.

[52] S. A. Meynell, M. N. Wilson, K. L. Krycka, B. J. Kirby, H. Fritzsche, T. L. Monchesky, *Phys. Rev. B* **2017**, *96*, 054402.

[53] J. López-López, J. M. Gomez-Perez, A. Álvarez, H. B. Vasili, A. C. Komarek, L. E. Hueso, F. Casanova, S. Blanco-Canosa, *Phys. Rev. B* **2019**, *99*, 144427.

[54] S. G. Azatyan, O. A. Utas, N. V. Denisov, A. V. Zotov, A. A. Saranin, *Surf. Sci.* **2011**, *605*, 289.

[55] S. Kahwaji, R. A. Gordon, E. D. Crozier, T. L. Monchesky, *Phys. Rev. B* **2012**, *85*, 014405.

[56] K. Schwinge, C. Müller, A. Mogilatenko, J. J. Paggel, P. Fumagalli, *J. Appl. Phys.* **2005**, *97*, 103913.

[57] S. Higashi, P. Kocán, H. Tochihara, *Phys. Rev. B* **2009**, *79*, 205312.

[58] Z.-Q. Zou, W.-C. Li, *Phys. Lett. A* **2011**, *375*, 849.

[59] C. Gong, L. Li, Z. Li, H. Ji, A. Stern, Y. Xia, T. Cao, W. Bao, C. Wang, Y. Wang, Z. Q. Qiu, R. J. Cava, S. G. Louie, J. Xia, X. Zhang, *Nature* **2017**, 546, 265.

[60] P. A. Joy, P. S. Anil Kumar, S. K. Date, *J. Phys.: Condens. Matter* **1998**, 10, 11049.

[61] P. Bruno, *Mat. Res. Soc. Symp. Proc.* **1992**, 231, 299.

[62] A. M. Tokmachev, D. V. Averyanov, A. N. Taldenkov, I. S. Sokolov, I. A. Karateev, O. E. Parfenov, V. G. Storchak, *ACS Nano* **2021**, 15, 12034.

[63] R. Roemer, D. H. D. Lee, S. Smit, X. Zhang, S. Godin, V. Hamza, T. Jian, T. Larkin, H. Shin, C. Liu, M. Michiardi, G. Levy, Z. Zhang, R. J. Green, C. Kim, D. Muller, A. Damascelli, M. J. Han, K. Zou, *npj 2D Mater. Appl.* **2024**, *8*, 63.

[64] D.-Y. Wang, X. Yang, W. He, Q.-F. Zhan, H.-F. Du, H.-L. Liu, X.-Q. Zhang, Z.-H. Cheng, *J. Magn. Magn. Mater.* **2021**, 538, 168252.

[65] N. Steinki, D. Schroeter, N. Wächter, D. Menzel, H. W. Schumacher, I. Sheikin, S. Süllow, *J. Phys. Commun.* **2020**, 4, 035008.

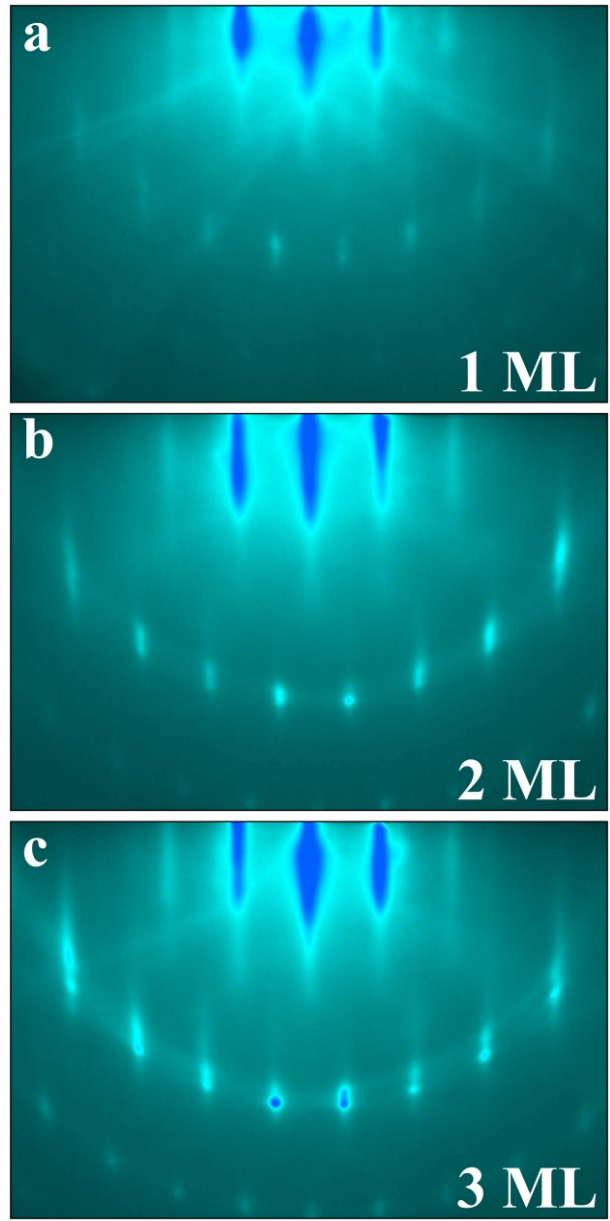

**Figure 1.** Characteristic RHEED patterns (along the $[1\bar{1}0]$ azimuth of the Si substrate) of a) 1 ML, b) 2 ML, and c) 3 ML MnSi films.

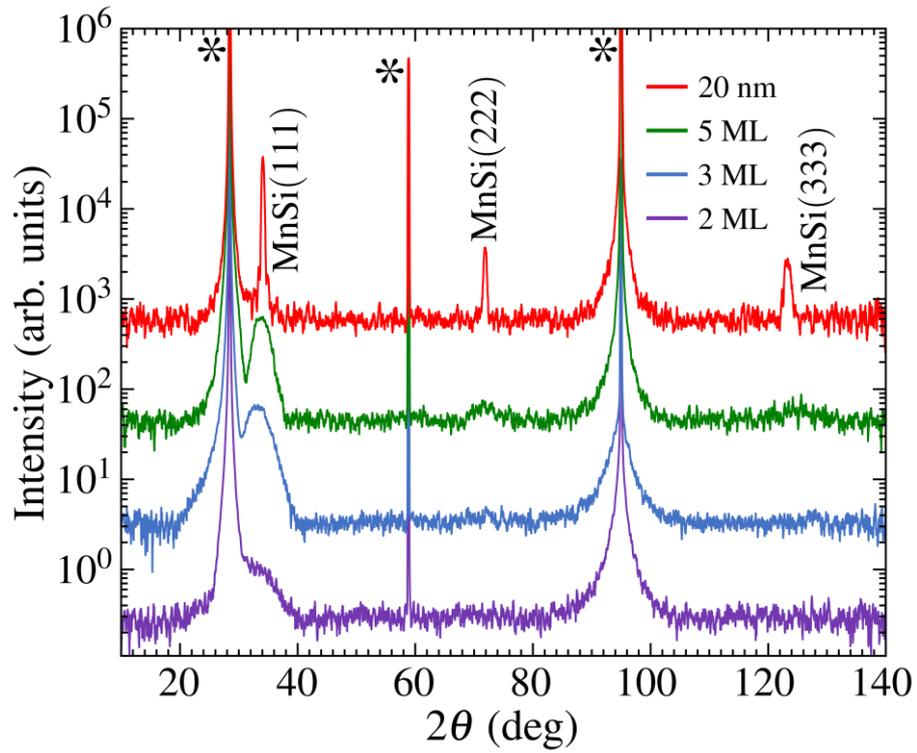

**Figure 2.** θ-2θ XRD scans of MnSi films: 2 ML (purple), 3 ML (blue), 5 ML (green), and 20 nm (red). To increase the visibility of the peaks, the scans are shifted vertically; the asterisks mark peaks from the Si(111) substrate.

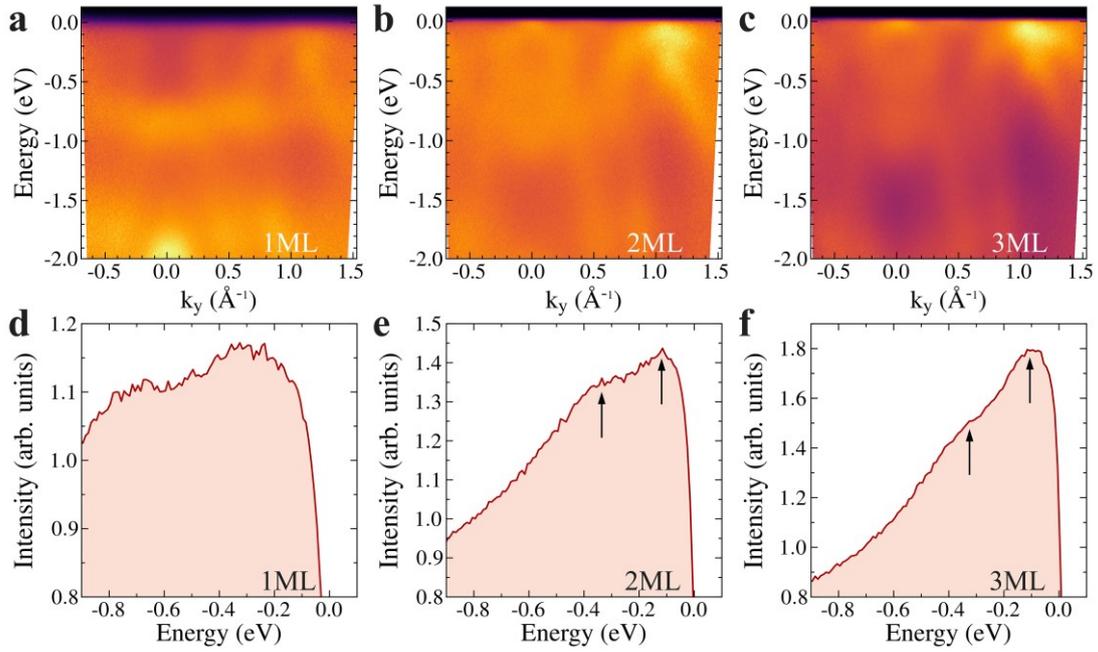

**Figure 3.** ARPES spectra of MnSi films of different thickness: a) 1 ML, b) 2 ML, and c) 3 ML. The hole-like band at -2 eV near $\overline{\Gamma}$ is from the Si(111) substrate, which is partially uncovered for 1 ML MnSi film due to slight inhomogeneity. EDCs at $k_y \sim 1.12$ Å for MnSi films of different thickness: d) 1 ML, e) 2 ML, and f) 3 ML. Arrows in (e,f) indicate split peaks from magnetic exchange coupling.

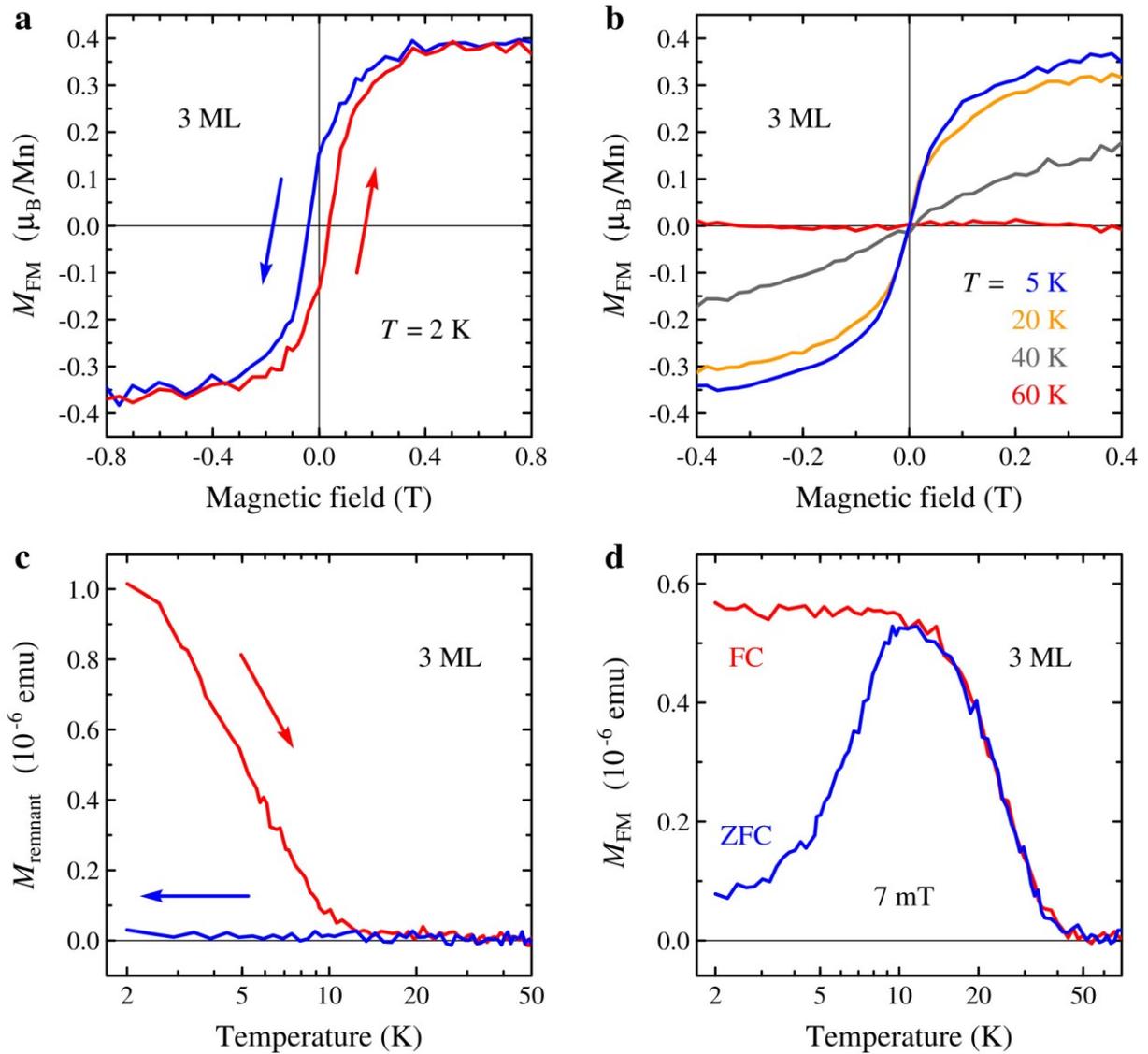

**Figure 4.** Magnetic properties of 3 ML MnSi. a) *M-H* hysteresis loop measured in in-plane magnetic fields at 2 K. b) Dependence of the magnetic moment on in-plane magnetic fields at 5 K (blue), 20 K (orange), 40 K (gray), and 60 K (red). c) Temperature dependence of the remnant magnetic moment after cooling in an in-plane magnetic field 1 T. d) Temperature dependence of the FM moment for zero-field cooling (ZFC, blue) and field cooling (FC, red) in an in-plane magnetic field 7 mT.

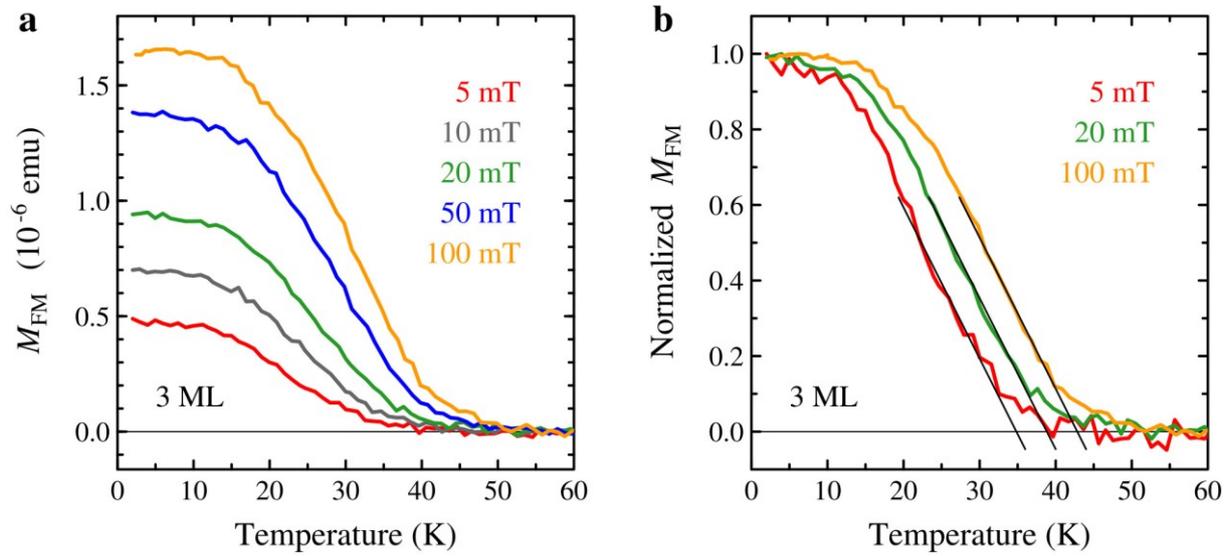

**Figure 5.** 2D nature of magnetism in 3 ML MnSi. a) Temperature dependence of the FM moment in in-plane magnetic fields 5 mT (red), 10 mT (gray), 20 mT (green), 50 mT (blue), and 100 mT (orange). b) Temperature dependence of the normalized FM moment in in-plane magnetic fields 5 mT (red), 20 mT (green), and 100 mT (orange).

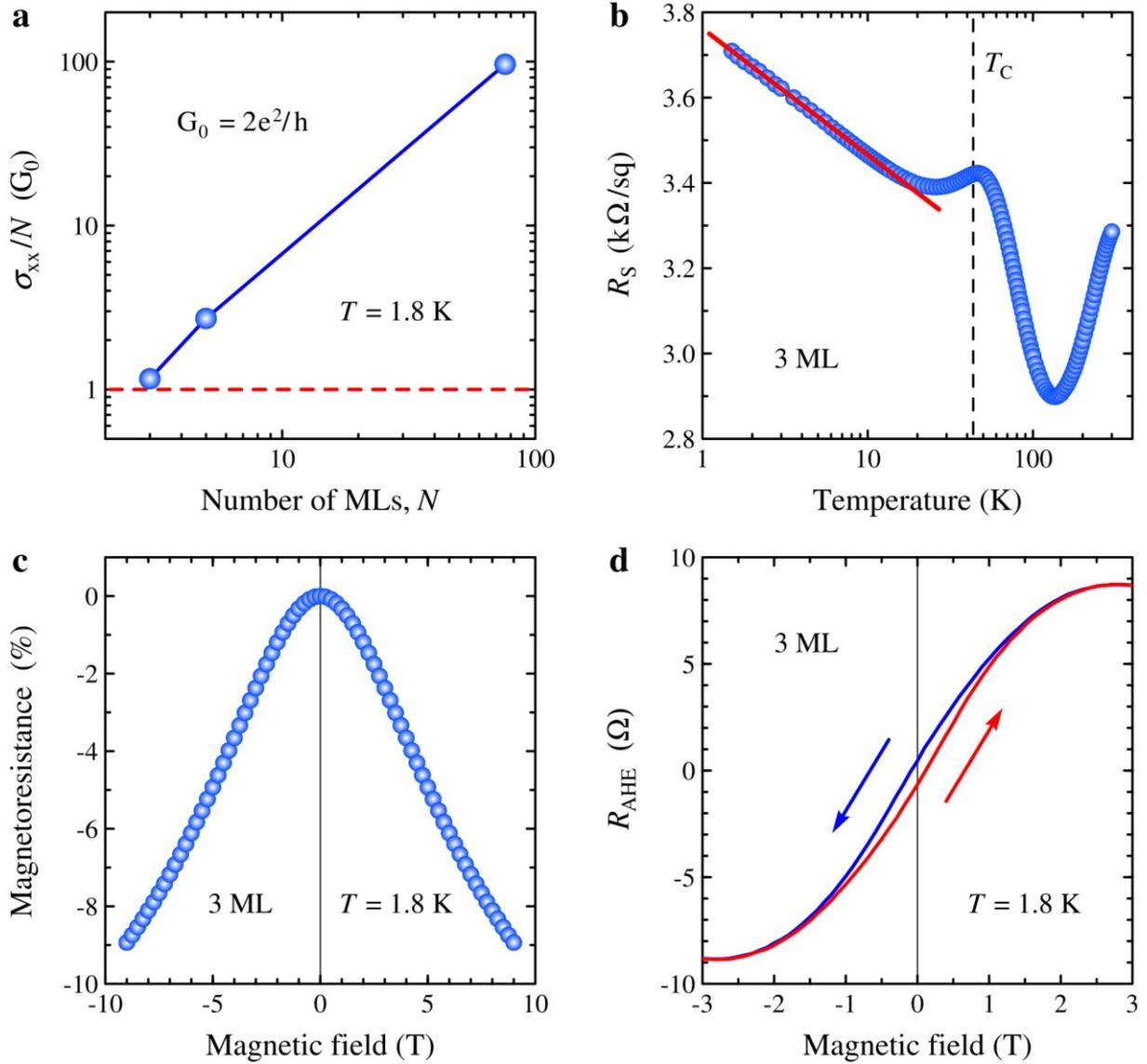

**Figure 6.** a) Dependence of layer normalized sheet conductance in MnSi films at 1.8 K on the number of MLs $N$. b) Temperature dependence of sheet resistance in 3 ML MnSi (blue dots). The red line highlights the logarithmic functional form of the dependence at low temperature. c) Magnetic field dependence of MR in 3 ML MnSi at 1.8 K. d) Hysteretic magnetic field dependence of the anomalous Hall effect resistance in 3 ML MnSi at 1.8 K.

# Supporting Information

**Two-Dimensional Ferromagnetism in Monolayers of MnSi**

*Yuan Fang, Yang Liu, Dmitry V. Averyanov, Ivan S. Sokolov, Alexander N. Taldenkov, Oleg E. Parfenov, Oleg A. Kondratev, Andrey M. Tokmachev, Vyacheslav G. Storchak\**

Content:

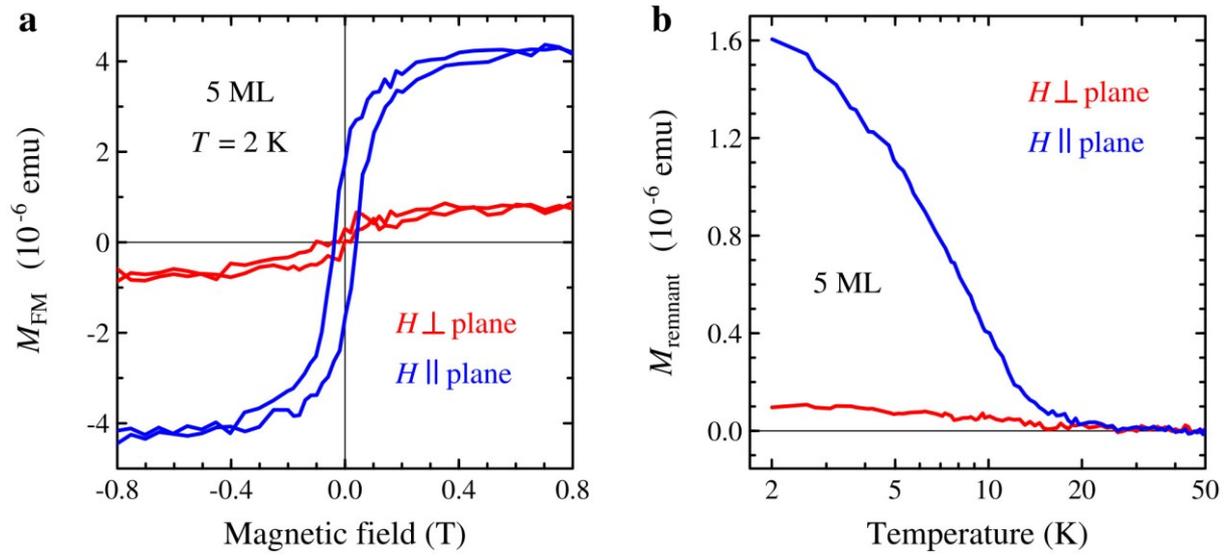

**Figure S1.** Magnetic anisotropy in 5 ML MnSi. a) *M-H* hysteresis loops measured in in-plane (blue) and out-of-plane (red) magnetic fields at 2 K. b) Temperature dependence of the remnant magnetic moment after cooling in an in-plane (blue) or out-of-plane (red) magnetic field 1 T.

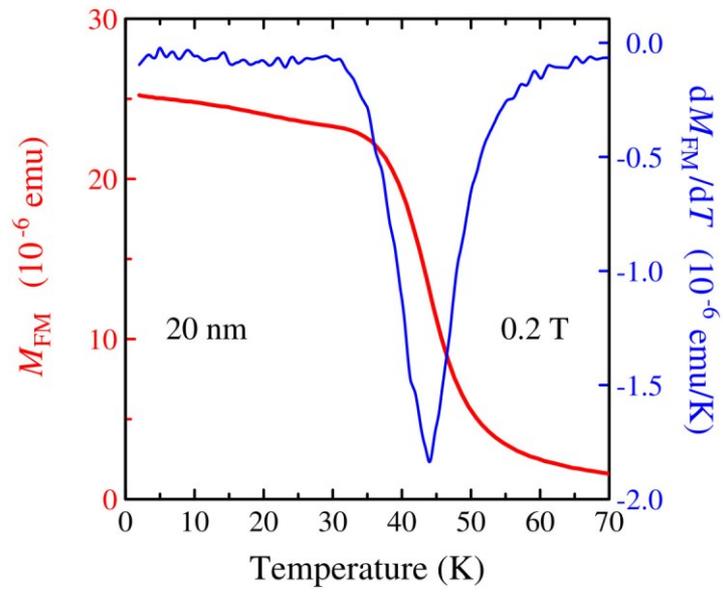

**Figure S2.** Temperature dependence of the FM moment in 20 nm MnSi measured in an in-plane magnetic field 0.2 T (red) and its derivative with respect to temperature (blue).

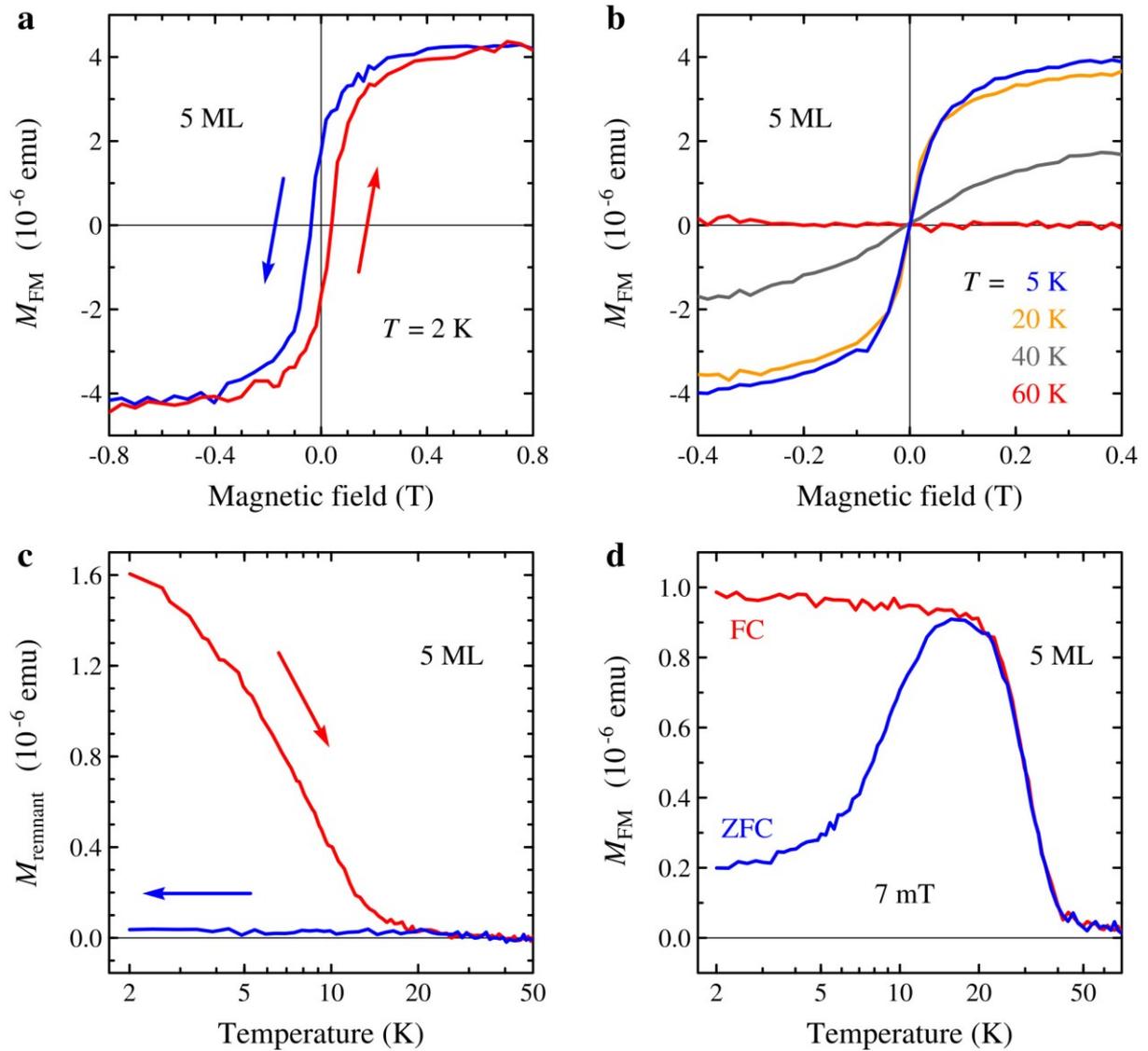

**Figure S3.** Ferromagnetic properties of 5 ML MnSi. a) *M-H* hysteresis loop measured in in-plane magnetic fields at 2 K. b) Dependence of the magnetic moment on in-plane magnetic fields at 5 K (blue), 20 K (orange), 40 K (gray), and 60 K (red). c) Temperature dependence of the remnant magnetic moment after cooling in an in-plane magnetic field 1 T. d) Temperature dependence of the FM moment for zero-field cooling (ZFC, blue) and field cooling (FC, red) in an in-plane magnetic field 7 mT.

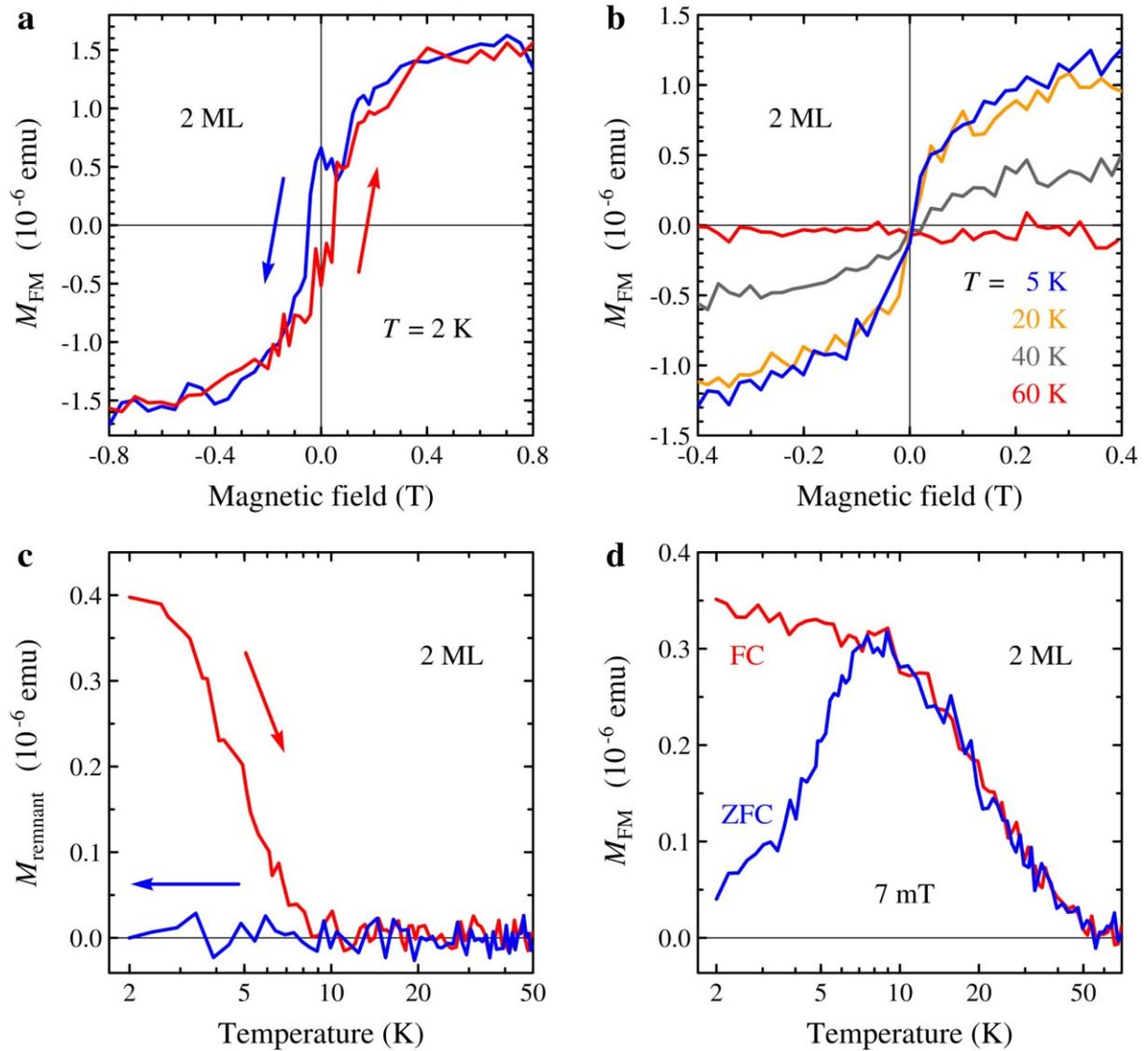

**Figure S4.** Ferromagnetic properties of 2 ML MnSi. a) *M-H* hysteresis loop measured in in-plane magnetic fields at 2 K. b) Dependence of the magnetic moment on in-plane magnetic fields at 5 K (blue), 20 K (orange), 40 K (gray), and 60 K (red). c) Temperature dependence of the remnant magnetic moment after cooling in an in-plane magnetic field 1 T. d) Temperature dependence of the FM moment for zero-field cooling (ZFC, blue) and field cooling (FC, red) in an in-plane magnetic field 7 mT.

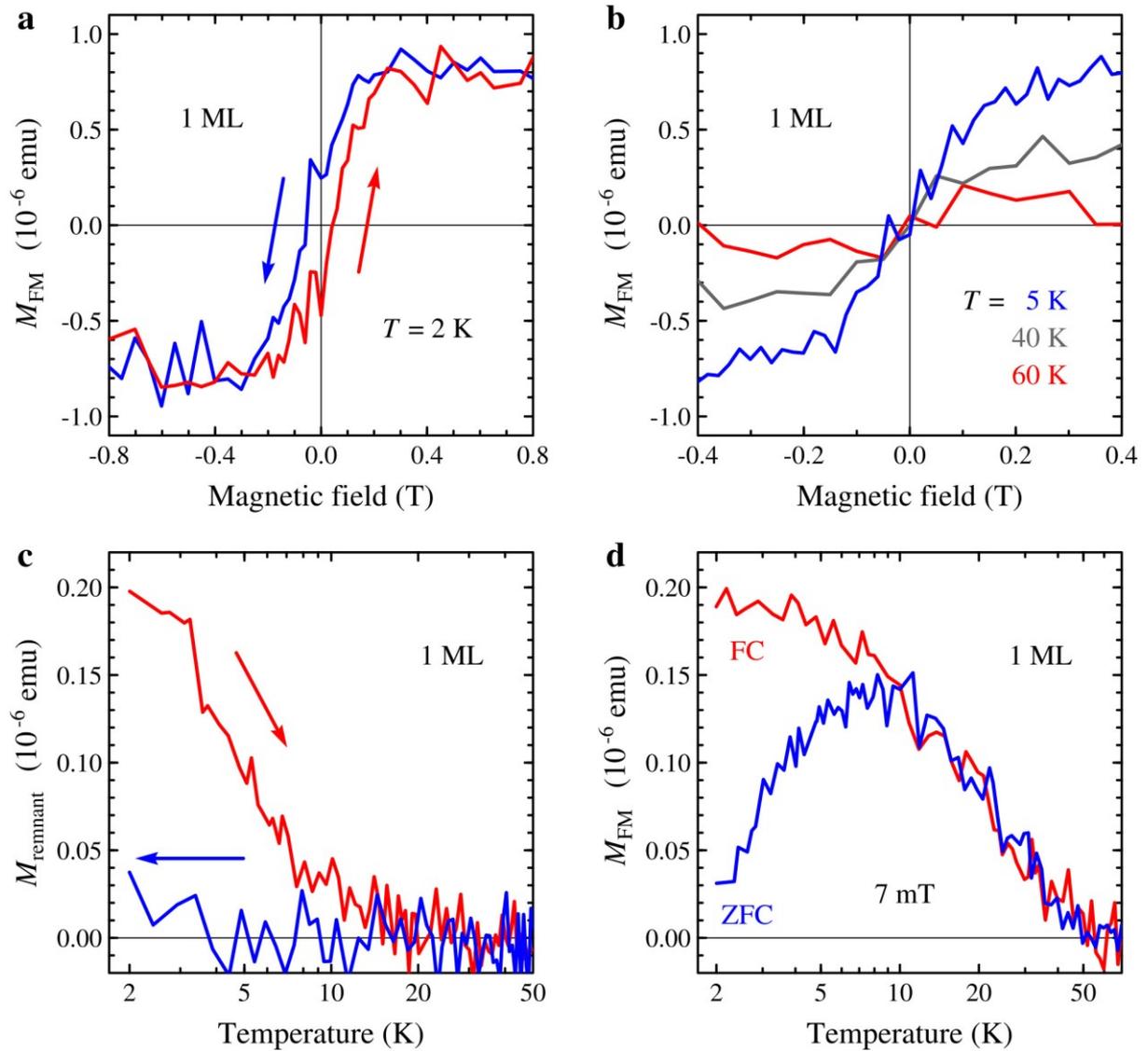

**Figure S5.** Ferromagnetic properties of 1 ML MnSi. a) *M-H* hysteresis loop measured in in-plane magnetic fields at 2 K. b) Dependence of the magnetic moment on in-plane magnetic fields at 5 K (blue), 40 K (gray), and 60 K (red). c) Temperature dependence of the remnant magnetic moment after cooling in an in-plane magnetic field 1 T. d) Temperature dependence of the FM moment for zero-field cooling (ZFC, blue) and field cooling (FC, red) in an in-plane magnetic field 7 mT.

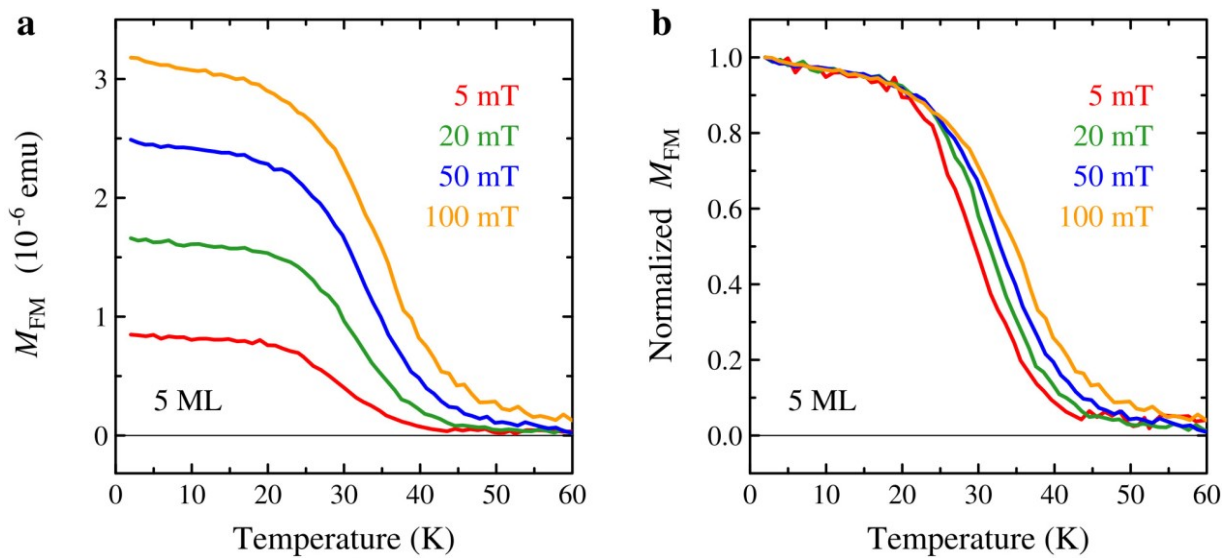

**Figure S6.** a) Temperature dependence of the FM moment in 5 ML MnSi in in-plane magnetic fields 5 mT (red), 20 mT (green), 50 mT (blue), and 100 mT (orange). b) Temperature dependence of the normalized FM moment in 5 ML MnSi in in-plane magnetic fields 5 mT (red), 20 mT (green), 50 mT (blue), and 100 mT (orange).

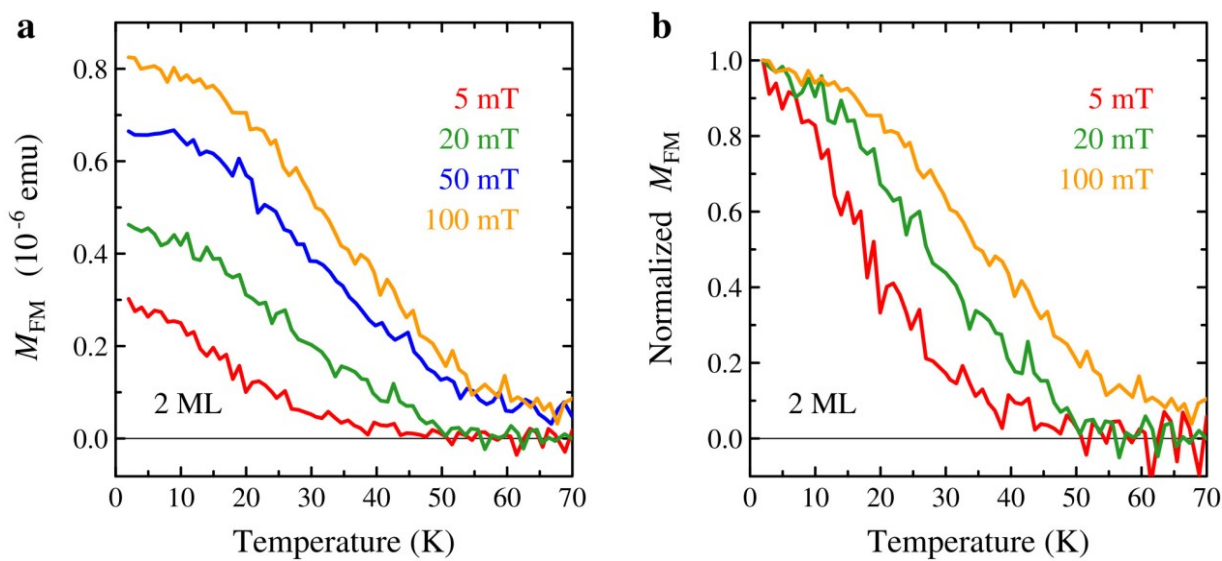

**Figure S7.** a) Temperature dependence of the FM moment in 2 ML MnSi in in-plane magnetic fields 5 mT (red), 20 mT (green), 50 mT (blue), and 100 mT (orange). b) Temperature dependence of the normalized FM moment in 2 ML MnSi in in-plane magnetic fields 5 mT (red), 20 mT (green), and 100 mT (orange).

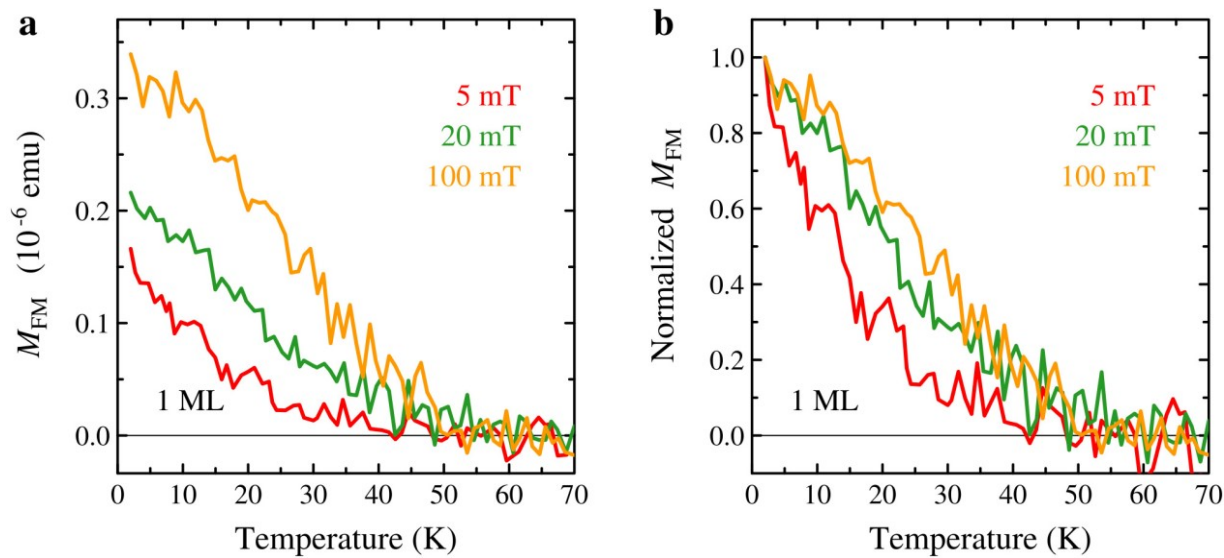

**Figure S8.** a) Temperature dependence of the FM moment in 1 ML MnSi in in-plane magnetic fields 5 mT (red), 20 mT (green), and 100 mT (orange). b) Temperature dependence of the normalized FM moment in 1 ML MnSi in in-plane magnetic fields 5 mT (red), 20 mT (green), and 100 mT (orange).

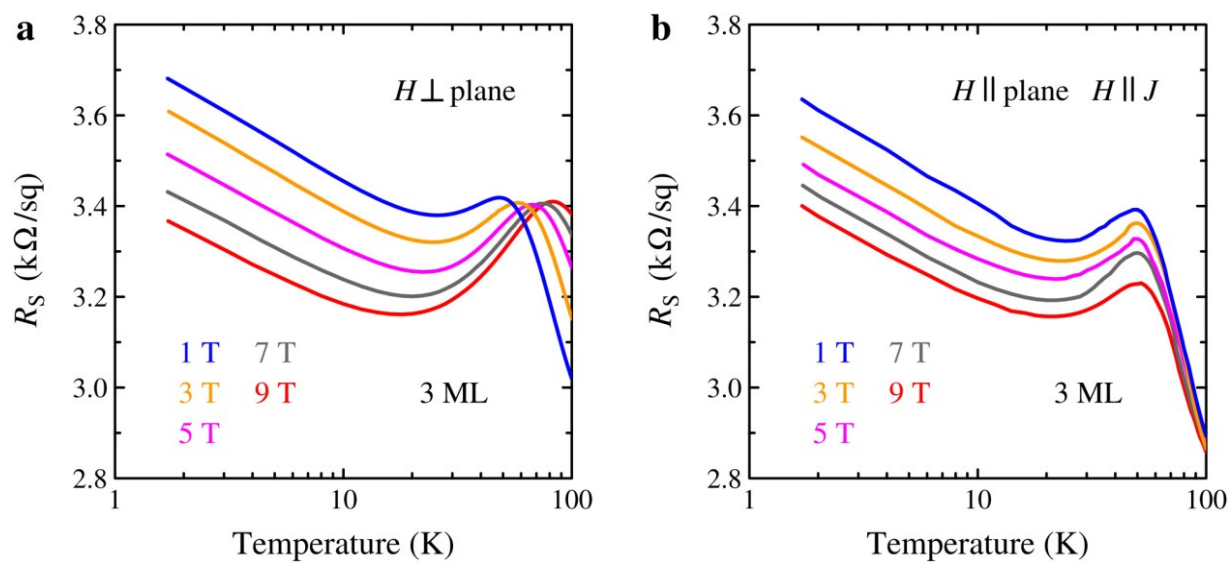

**Figure S9.** Temperature dependence of sheet resistance in 3 ML MnSi in magnetic fields 1 T (blue), 3 T (orange), 5 T (magenta), 7 T (gray), and 9 T (red) directed a) out-of-plane and b) in-plane parallel to the current.

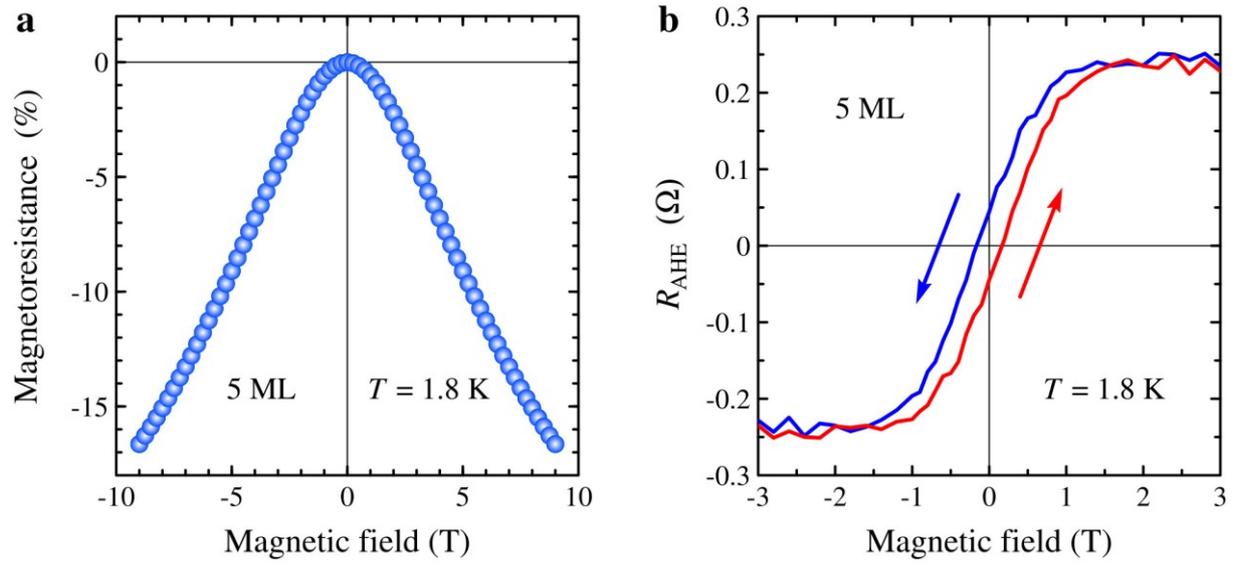

**Figure S10.** a) Magnetic field dependence of MR in 5 ML MnSi at 1.8 K. b) Hysteretic magnetic field dependence of the anomalous Hall effect resistance in 5 ML MnSi at 1.8 K.